# Knots and Divergences

Dirk Kreimer
Dept. of Physics
University of Tasmania
GPO Box 252C
Hobart
TAS 7001
Australia

March 10, 1995

**Abstract**

We give empirical evidence that the UV-divergences of a renormalizable field theory are knot invariants.

## 1 Introduction

Since a long time some number-theoretic results in perturbative calculations in quantum field theories have presented a challenge to theorists. Lot of authors who published results of multiloop calculations suspected a hidden and yet to be understood structure governing the rational and transcendental numbers which arise as the coefficients describing the UV-divergent structures of the theory [1, 2]. Especially in dimensional regularization, the cleanest bookkeeping method for UV-divergences we have invented so far, the numbers arising in MS $Z$-factors for example are highly suspicious. It is the purpose of this letter to argue that these numbers can be understood as knot invariants, that is containing topological information identfying knots to be associated with the Feynman graphs in some prescribed manner. It is not the purpose of the paper to derive this connection from first principles, but we rather collect empirical evidence to support our case.

So we want to assign certain topological properties of a Feynman graph to some number-theoretic properties of its value when calculated in dimensional regularization.

We will split our reasoning into two parts. First we argue that the topologically simplest Feynman graphs -the ladder topologies- have UV-divergences which are free of transcendental numbers. This is a necessary condition to give meaning to our second, empirical, finding: The transcendental coefficients arising in the UV divergent part of more complicated diagrams describe knots in these diagrams.

In the next section we discuss topologically simple Feynman graphs by introducing certain algebraic structures valid for ladder topologies. In the following section, we descibe how to assign a link diagram to a Feynman graph. In the core section of the paper we compare known results for various Feynman graphs with the knots identified in them. In fact, we identify the Feynman graphs which generate the $(2,q)$ torus knots, and will report on the $(3,4)$ torus knot which appears at the six-loop level.

This letter is mainly based on a previous paper [3], where especially the results of the next section are given in much more detail and generality.

## 2  Algebraic Structures in One-loop Integrals

Here we want to discuss Feynman graphs which have a simple ladder topology, cf. Fig.(1). We report here on some findings in [3], where the reader will find details.

It is a remarkable consequence of renormalization theory that for ladder topologies, the whole renormalization program can be absorbed in a simple one-loop algebra. We describe this algebra in the example of a three-point function at momentum transfer zero. The crucial point is to utilize the fact that overall divergences are independent of internal masses and momentum transfer. This allows us to consider massless three-point functions, where it is understood that the subdivergences have to be taken into account appropriately.

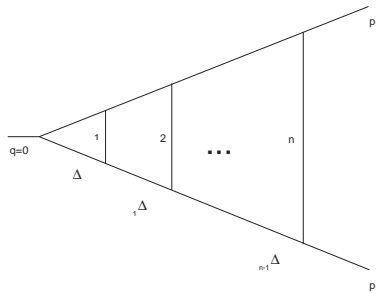

Fig.(1) By evaluating at zero momentum transfer the calculation of the massless ladder $\bar{\Gamma}^{(n)}$ becomes a concatenation of $_j\Delta^1$ functions.

Take as an example two-loop scalar QED. We get

$$\bar{\Gamma}^{(2)} = {}_0\Delta^1 \, {}_1\Delta^1 \, (q^2)^{-2\varepsilon} \, \Gamma^{(0)}, \tag{1}$$

where we define ${}_j\Delta^1$ by

$$\int d^D k \frac{(k^2)^{-\varepsilon j}}{\not{k}\not{k}(k+p)^2} = \int d^D k \frac{(k^2)^{-\varepsilon j}}{k^2(k+p)^2} =: {}_j\Delta^1 \, (q^2)^{-\varepsilon(j+1)} \, \Gamma^{(0)}. \tag{2}$$

We consider the functions ${}_j\Delta^1$ as modified one-loop functions. For any renormalized theory they can be obtained from the corresponding standard one-loop integral by a change in the measure

$$\int d^D k \to \int d^D k (k^2)^{-\varepsilon j}.$$

We will see that the index $j$ can be identified as a writhe number later on.

Eventually, we end up with a product of concatenated generalized one-loop functions for the $n$-loop diagram, Fig.(1). Introducing a projector $< \ldots >$ onto the UV-divergences (the proper singular part of a Laurent expansion in $\varepsilon$, where $\varepsilon$ is the DR regularization parameter) so that, by definition,

$$< \text{UV-finite expression} > = 0, \tag{3}$$

we can summarize the whole result for the graph and its counterterms in the following way:

We define two operators

$$\begin{aligned}
B^k(\Delta^1) &:= \prod_{i=0}^{k} {}_i\Delta^1, \\
A^r(\alpha) &:= \Delta^1 <<< \ldots < \alpha > \Delta^1 > \ldots \Delta^1 >, \\
& \quad \text{r} <> \text{brackets} \\
\Rightarrow A^r(B^k(\Delta^1)) &= \Delta^1 <<<< \ldots < \prod_{i=0}^{k} {}_i\Delta^1 > \Delta^1 > \ldots \Delta^1 >, \\
B^k(A^r(\Delta^1)) &= <<< \ldots < \Delta^1 > \Delta^1 > \ldots \Delta^1 > \prod_{i=0}^{k} {}_i\Delta^1, \\
B^0(\Delta^1) = A^0(\Delta^1) &= \Delta^1.
\end{aligned} \tag{4}$$

$B$ acts by concatenating massless one-loop functions with increasing writhe number, $A$ by projecting on the divergent part of products iteratively, thus taking into account subdivergences.

The general result for $Z_1^{(n)}$, which we define to be the MS-$Z$-factor, calculated for the graphs of Fig.(1) only, follows immediately:

$$Z_1^{(n)} = < [-A + B]^{(n-1)}(\Delta^1) > . \tag{5}$$

For our two-loop example this delivers
$$Z_1^{(2)} = <\Delta^1 {}_1\Delta^1 - <\Delta^1> \Delta^1>, \tag{6}$$
and for $n = 3$ it gives
$$Z_1^{(3)} = <\Delta^1 {}_1\Delta^1 {}_2\Delta^1 - (<\Delta^1 {}_1\Delta^1> - <<\Delta^1>\Delta^1>)\Delta^1 \\ - <\Delta^1>\Delta^1 {}_1\Delta^1>. \tag{7}$$

In [3] it is proven that such an algebraic structure also persists in the case of two-point functions, and overlapping instead of nested divergences. In fact, it exists whenever we have a simple ladder topology.

More striking, it was shown in [3], that in the sum of the graph together with its counterterm contributions all transcendental coefficients in the divergences drop out. As an example we give explicit expressions for ${}_3\Delta$ and for $Z^{(3)}$. We do not use a renormalization which would absorb the $\gamma$ and $\zeta(2)$, as we want to exhibit the generated rationals in their purest form.

$$\begin{aligned}
Z^{(3)} &= \frac{5}{4}\frac{1}{(D-4)} - \frac{19}{24}\frac{1}{(D-4)^2} + \frac{1}{4}\frac{1}{(D-4)^3} - \frac{1}{24}\frac{1}{(D-4)^4}, \\
\bar{\Gamma}^{(3)} &= \frac{1}{(D-4)}\left(-\frac{23}{9}\zeta(3) + \frac{1}{3}\zeta(2)\gamma - \frac{7}{6}\zeta(2) - \frac{4}{9}\gamma^3 + \frac{14}{3}\gamma^2 - \frac{125}{6}\gamma + \frac{455}{12}\right) \\
&+ \frac{1}{(D-4)^2}\left(\frac{1}{12}\zeta(2) + \frac{1}{3}\gamma^2 - \frac{7}{3}\gamma + \frac{125}{24}\right) \\
&+ \frac{1}{(D-4)^3}\left(-\frac{1}{6}\gamma + \frac{7}{12}\right) \\
&+ \frac{1}{(D-4)^4}\left(\frac{1}{24}\right).
\end{aligned}$$

In general, the cancellation can be proved using combinatoric properties of the function
$$_j\Delta = \frac{\Gamma(1+(j+1)\varepsilon)\Gamma(1-\varepsilon)\Gamma(1-(j+1)\varepsilon)}{(j+1)\varepsilon(1-(j+2)\varepsilon)\Gamma(1+j\varepsilon)\Gamma(1-(j+2)\varepsilon)}.$$

Define
$$\begin{aligned}
P_n &:= \prod_{i=0}^{n-1} {}_i\Delta, \\
\Rightarrow P_n &= \frac{(\Gamma(1-\varepsilon))^{n+1}\Gamma(1+n\varepsilon)}{n!\varepsilon^n(1-2\varepsilon)\ldots(1-(n+1)\varepsilon)\Gamma(1-(n+1)\varepsilon)}.
\end{aligned}$$

Now use
$$\Gamma(1-z) = \exp(\gamma z)\exp\left(\sum_{j=2}^{\infty}\frac{\zeta(j)}{j}z^j\right), \tag{8}$$

to obtain

$$P_n = \frac{1}{n!\varepsilon^n(1-2\varepsilon)\ldots(1-(n+1)\varepsilon)}\exp(-n\gamma\varepsilon)$$
$$\exp(\sum_{j=2}^{\infty}\frac{\zeta(j)}{j}\varepsilon^j[n+1+(-n)^j-(n+1)^j]),$$

which is the starting point of the proof in [3].

## 3  From Feynman Graphs to Link Diagrams

We give here a simple prescription how to assign link diagrams to Feynman graphs. Assume we have drawn the Feynman graph in a way that all vertices are located on a circle (we could call this the Hamiltonian circuit representation). This will not work for the general case, but is sufficient for our purposes here. (In general, not every three-valent graph allows for a Hamiltonian circuit, but the failure appears at loop orders and topologies which are not relevant for our purposes.[4])

For a $n$-loop graph there are $n-1$ propagators not on this circle connecting internal vertices. Now replace every vertex by an over- or undercrossing according to the following rules.

- Every loop in the Feynman diagram corresponds to a link. Correspondingly, a $n$-loop diagram will map to a link diagram consisting of $n$ links.

- The links are oriented according to the flow of loop momenta, and follow the rule that at every vertex the momentum coming from the right is overcrossing as in Fig.(2):

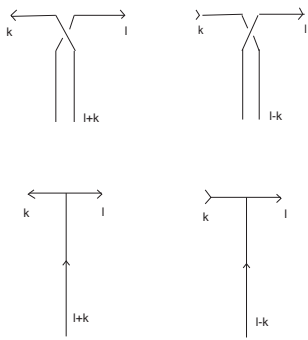

Fig.(2) The replacement of a three-point vertex by an overcrossing. When we reverse the orientation of lines at the vertex, we have also to reverse the orientation of lines in the link diagram, and, accordingly, exchange the over- to an undercrossing.

Also propagators which cross each other in the Feynman graph due to its topological nature will follow the *from the right = overcrossing* rule.

The only crossings we allow in the link diagram correspond to vertices or crossings of propagators in the Feynman graph. Now let us study the ladder topologies first.

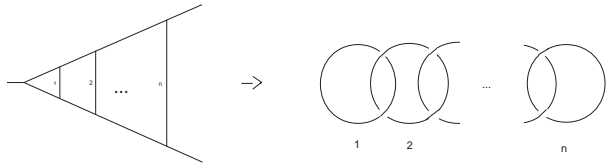

Fig.(3) The translation from a Feynman diagram into a link diagram. Each vertex is replaced by an over/undercrossing according to the momentum flow at the vertex. We follow the convention to have the momentum flow in each loop counterclockwise. Here we used a momentum routing so that each propagator $P_i, i = 1, \ldots, n$, appearing as a rung in the Feynman graph above, carries loop momentum $l_i - l_{i+1}$.

For the ladder topology we understand that each crossing in our link diagram has to correspond to a vertex in the Feynman graph. If we were to choose other momentum routings we might generate other link diagrams, for example:

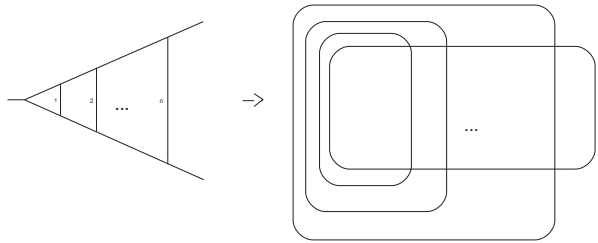

Fig.(4) A different routing of momenta. This time we have chosen the propagators $P_i$ to carry loop momentum $l_i$ and demand that each loop momentum passes the exterior vertex on the left.

Now apply a skein relation (Fig.(5)) to the link diagram. The idea is that with the help of a skein relation we can transform our $n$-component link diagram to a one-component knot. For this to be achieved we have to apply the skein relations $n - 1$ times. We apply the skein relation only to crossings which correspond to vertices in the graph. We exclude its application to crossings which correspond to mutual crossings of propagators in the Feynman graph, as such crossings can be arbitrarily generated by drawing a Feynman graph in various different ways.

Returning to our ladder topology, we see that in both cases, Fig.(3) and Fig.(4), applying the skein relation $n-1$ times, we find the same result: Each application of the $X$ part disentangles a ring, while the $Y$ part concatenates rings together, creating terms with non-vanishing writhe number; in particular the term of degree $Y^{n-1}$ is an unknot with writhe $n-1$. Fig.(6) is an example for the two-loop case.

$$X \;\times\; + \;Y\; || \;=\; \times$$

Fig.(5) The skein relation, an exchange identity which allows the disentangling of the link diagram. $X$ and $Y$ have to be regarded as operators to be identified with $A$ and $B$ in an appropriate manner, see Fig.(6).

Fig.(6) An explicit two-loop example. The last line indicates how the operators $X, Y$ have to be identified. Observe that the $X$ part disentangles the rings, while the $Y$ part concatenates them to an unknot with writhe 1.

We see that with the identifications

$$X^{r-1}(\bigcirc \bigcirc \ldots \bigcirc) \;\Rightarrow\; [-A]^{r-1}(\Delta^1),$$
$$Y^{r-1}(\overbrace{\bigcirc\!\bigcirc \ldots \bigcirc\!\bigcirc}_{r-1}) \;\Rightarrow\; B^{r-1}(\Delta^1),$$

we obtain our previous result. So we have identified the unknot with an appropriate one-loop function $_0\Delta^1$ and links of the form of Fig.(7)

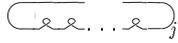

Fig.(7) A link with writhe number $j$.

with the corresponding function $_j\Delta^1$. Note that this implies that we have no invariance under Reidemeister type $I$ moves, so that we work with a regular isotopy.

Let us consider some more complicated topologies next. The first candidates for a non-ladder topology are the following three-loop graphs, given in Fig.(8) with the corresponding link diagram Fig.(9).

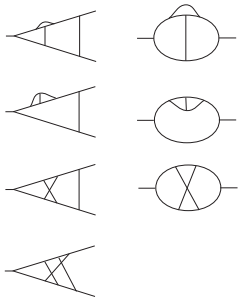

Fig.(8) Topological non-simple three-loop graphs. They all involve $\zeta(3)$ in their divergent part, even after adding counterterm contributions.

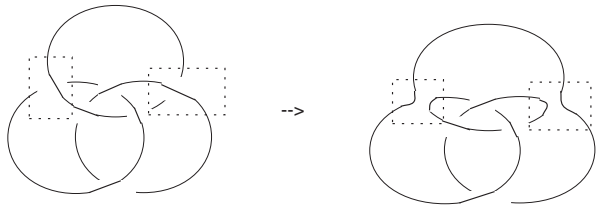

Fig.(9) The corresponding topology. The dashed rectangles indicate where the $Y$ part of the skein relation has been applied twice.

Assume we remove one of the internal propagators in the graphs of Fig.(8). We then have a two-loop ladder topology, corresponding to a two-component link diagram. Reinserting the propagator corrresponds to entangling a third link of this link diagram. Due to the fact that this propagator actually crosses the other one, therefore distinguishing this diagram from the three-loop ladder diagram, the reader can easily convince himself that the third ring will generate four more crossings in the link diagram, no matter what the momentum routing in the Feynman diagram was. So we end up with the six crossing diagram on the lhs of Fig.(9). For example using our standard assignment of loop momenta, two of the new crossings correspond to the two vertices we used for the propagator to be reattached, and two more crossings stem from the crossings of this propagator with the other propagator: it carried two-loop momenta and therefore accounts for two lines in the link diagram. Other routings of loop momenta give the same result. We avoid any further crossings. We attach the non-planar propagator in the most economic way, by using the least number of crossings necessary to fulfill the rules above.

The link diagrams generated by these rules are of a very special kind: the crossings can be read off from the momentum flow, which we take to be counter-clockwise in all links. Thus in Fig.(10-18) we need not indicate the crossings by broken lines, in the manner of Fig.(9). The knot-theoretic consequence of this restricted class of link diagrams is that the knots they generate, by skeining, are closures of positive braids: a crucial restriction which reduces the number of possible knots with 8 crossings from 21 to 1.

Now note that by using a Reidemeister III move, we can transform the link diagram, and then use a Reidemeister I move to cancel a further crossing.

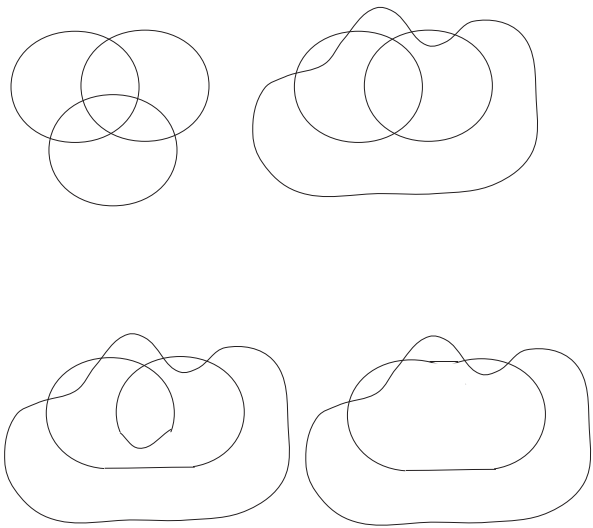

Fig.(10) Using Reidemeister II and III moves, we see that the crossings of ladder rungs are removable by Reidemeister I moves. We omit to draw over/undercrossings explicitly. They are determined by taking into account that all loops run counter-clockwise.

This is a general property: Assume we draw all propagators in the interior of the Hamiltonian circle. Now let us remove as many propagators as necessary to make the diagram planar (a ladder topology, that is). Using our standard momentum routing for this reduced diagram, let us begin to attach the non-planar propagators again, this time always using Reidemeister moves to avoid crossings with the rungs of the ladder.

Skeining the $r$ components of the ladder, it is clear that we get $r - 1$ Reidemeister I moves for free, removing $r - 1$ crossings in the link diagram. There remains a link diagram with $n - r$ components, which still has to be skeined.

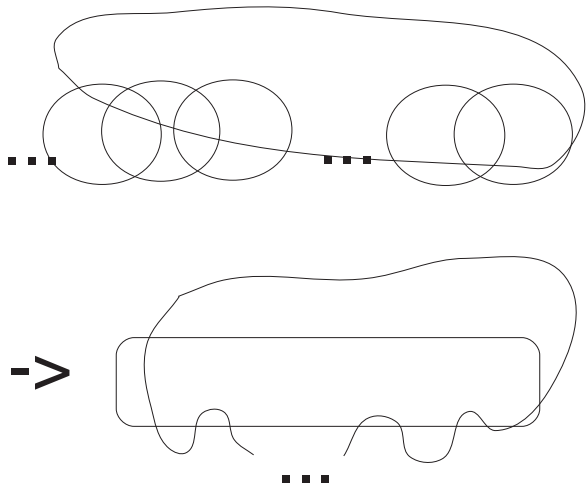

Fig.(11) The same as in Fig.(10), for the general case. The square dots here and in the following Fig.(12-14) indicate a continuation to arbitrary loop numbers.

We learn that as far as the knot content of the link diagram is concerned, the planar rungs are irrelevant. In the next section we will become familiar with the assignments of link diagrams to Feynman graphs by studying various examples.

## 4 Empirical Evidence for Knots

In section 1 we learned that ladder topologies are free of transcendentals. In section 2 we gained first experience with the translation of a Feynman diagram to a link diagram. In this section we try to relate transcendental numbers in the divergent part of Feynman diagrams to knots identified in them when skeining the associated link diagram. For simple ladder topologies, we found a way to actually calculate the relevant graphs in terms of one-loop integrals, thereby assigning some operational meaning to the coefficients of the skein relation. Such an operational definition of the skein relation which would extract the divergent part even for arbitrary topologies, starting from one-loop functions, eluded us so far, but we are content to report on some empirical findings relating again topology to number-theoretic properties. In the following we restrict ourselves to graphs without subdivergences, but with a non-vanishing overall degree of divergence.

So we now simply start considering Feynman graphs, their associated link diagrams, and the transcendentals one is confronted with in their calculation.

Consider as a first class of graphs the ones of Fig.(12).

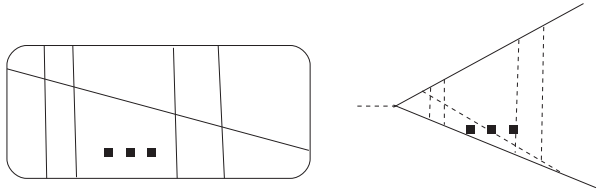

Fig.(12) These "slashed ladders" topologies generate the $(2, q)$ torus knots. On the right, we give an scalar QED example for Feynman graphs realizing this topology. Another example can be obtained in $\phi^4$ theory by considering the famous zig-zag topology [5].

It is a well-known fact [6] that these topologies have, in the $n$-loop case, the form

$$\frac{C_n \zeta(2n-3)}{(D-4)} + \mathcal{O}(1),$$

with rational $C_n$. Now we want to use our link approach to relate the transcendental $\zeta(2n-3)$ to knots. Therefore, we investigate this topology graphically in Fig.(13):

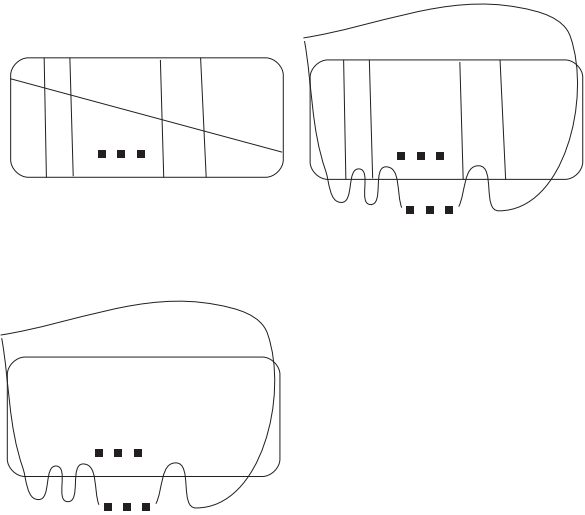

Fig.(13) Using the prescription of section 2 we find the braidword $\sigma^{2n-3}$, corresponding to the $(2, 2n-3)$ torus knot. Here again we used Reidemeister type I moves in the last step.

We conclude that $\zeta(2n-3)$ corresponds to the $(2, 2n-3)$ torus knot. For the case of the $\phi^4$ zig-zag topology, the rational $C_n$ could be determined for arbitrary $n$ and will be reported in [5].

It is interesting to obtain the same result using a different routing of momenta, Fig.(14).

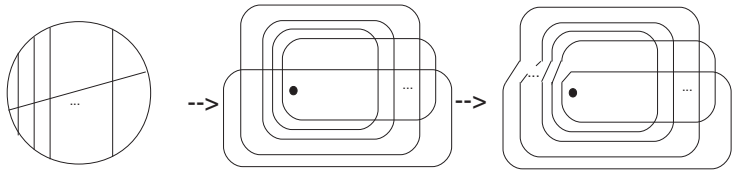

Fig.(14) The same with a different routing of momenta (for the ladder part we have chosen the momentum assignment of Fig.(4)). We omitted over/undercrossings in the above picture, as they are clear from the counterclockwise orientation of the loop momenta. Note that the link diagrams encircle the indicated dot counterclockwise, so that we can read off the corresponding braid words easily.

We read off from the above picture the braid group expression:

$$\sigma_{n-1} \ldots \sigma_1 \sigma_2 \ldots \sigma_{n-1} \sigma_1 \ldots \sigma_{n-2} = \sigma_{n-2}^{2n-3},$$

after applying Markov- and Reidemeister-moves. For example, chosing $n = 4$, we calculate

$$\sigma_3\sigma_2\sigma_1\sigma_2\sigma_3\sigma_1\sigma_2 = \sigma_2\sigma_1\sigma_2\sigma_1\sigma_3\sigma_2\sigma_3 =$$
$$\sigma_2\sigma_1\sigma_2\sigma_1\sigma_2\sigma_3\sigma_2 = \sigma_2\sigma_2\sigma_1\sigma_2\sigma_1\sigma_2\sigma_3 =$$
$$\sigma_2\sigma_2\sigma_1\sigma_2\sigma_1\sigma_2 = \sigma_2\sigma_2\sigma_2\sigma_1\sigma_2\sigma_2 =$$
$$\sigma_2^5\sigma_1 = \sigma_2^5.$$

So we have the beautiful correspondence $\zeta(2n-3)$ *in the Feynman graphs* ↔ $(2, 2n-3)$ *torus knot in the link diagram.*

Here we give two other typical results. First consider graphs which are two-particle reducible, as in Fig.(15):

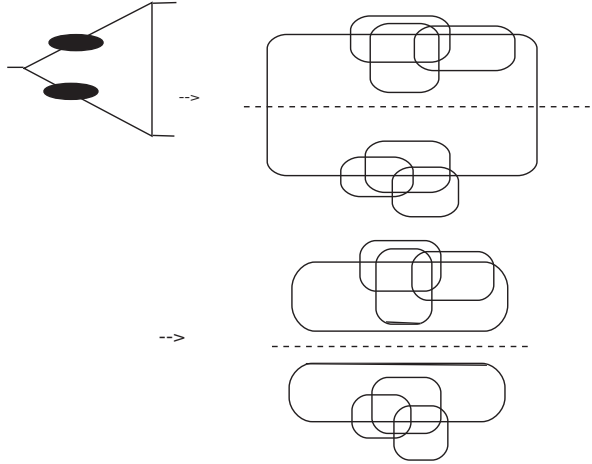

Fig.(15) Two-particle reducible graphs produce factor knots. The dashed line only cuts two lines and separates two independent knots.

These graphs produce independent maximal forests. The above picture shows that this produces link diagrams which are 2-line-reducible. This is the defining condition for a factor knot, so that the corresponding transcendentals for both factors correctly multiply.

Also relations between various graphs can be predicted, Fig.(16):

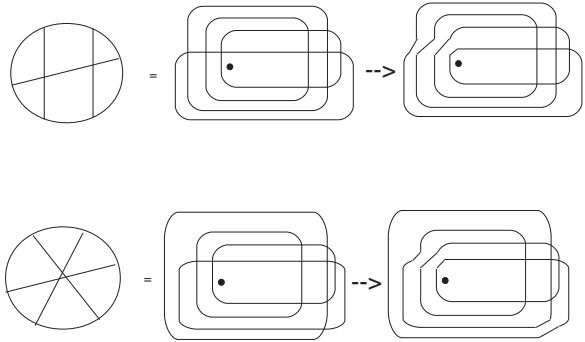

Fig.(16) These two Feynman graphs do not look the same, but both give $\zeta(5)$ [7]. It also follows immediately from the link diagram, by noticing that both generate the same knot.

Both give the $(2,5)$ torus knot as expected from our previous considerations.

But in multi-loop calculations one finds occasionally new transcendentals, independent from $\zeta(i)$. A first and prominent example is the transcendental found by David Broadhurst in a six-loop calculation at transcendentality level 8 [1]. According to our experience with $\zeta$-transcendentals, we would expect a knot with 8 crossings to appear in the corresponding Feynman graph.

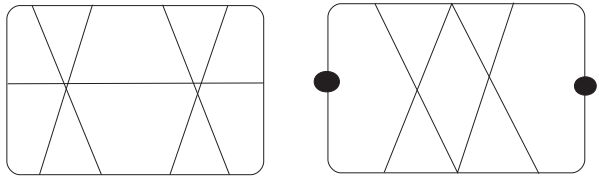

Fig.(17) We like to investigate this six-loop Feynman graph. We also give a $\phi^4$ graph which is topologically equivalent, and which was investigated by Broadhurst [1]. The two dots in this graph have to be identified. It can be obtained from the graph on the lhs by shrinking three propagators.

Let us map this graph to a link diagram, Fig.(18):

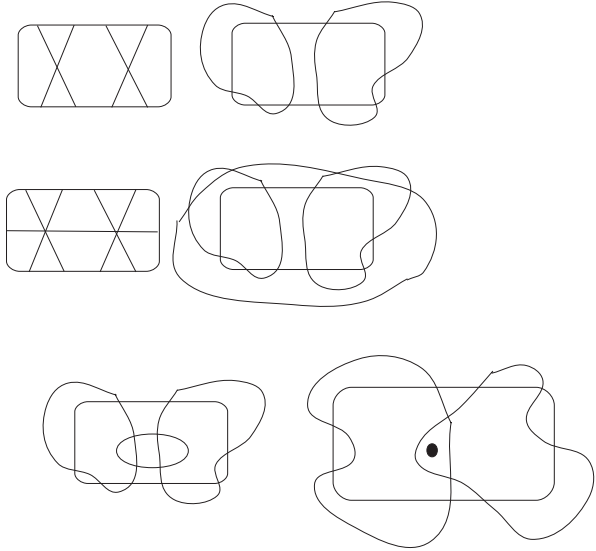

Fig.(18) The generation of the $(3,4)$ torus knot. In the first line we have removed one propagator to generate the $\zeta(3)\zeta(3)$ factor knot. Then we attach the last propagator in the most economic way, giving us the link diagram on the bottom rhs. We used Reidemeister II and III moves to get from the second to the third line. We end up with the braid word $\sigma_1^4\sigma_2\sigma_1^4\sigma_2$. (All components encircle the dot in the middle counterclockwise, so that we can read off the braid word.) After skeining the two kidneys we find a knot. It can be identified as the $8_{19}$ knot in the standard tables [8], which is the $(3,4)$ torus knot.

The identification of the $(3,4)$ torus knot was done by reading off the braid word
$$\sigma_1\sigma_2^3\sigma_1\sigma_2^3 = (\sigma_1\sigma_2)^4$$
from the above picture. In general, the $(p,q)$ torus knot has braid word [9]
$$(\sigma_1\ldots\sigma_{p-1})^q.$$
Thus the following picture emerges. According to the recipe outlined in section 2 we expect only positive braid words to appear. (Positive braid words have positive powers of braid generators only.) The positive braid words up to crossing number 9 are the $(2,q), q = 3, 5, 7, 9$ torus knots plus the $(3,4)$ torus knot. Crossing number 9 is the transcendentality level 9, which is exhausted by graphs up to six loops. An investigation of the results in [1] confirmed this pattern and will be reported in detail in [5]. There, we will also report on an investigation of the seven loop level, where some positive braid words appear which, for the first time, do not correspond to torus knots.

Further, as the $(3,4)$ torus knot is the only non-$\zeta$-ish transcendental at level 8, we conclude that our knot theoretic approach predicts a relation between the value of the transcendental $M$ which Broadhurst reported in [1] and the level 8 transcendental U62 which was found in the expansion of the master function [10]. And indeed, such a relation was meanwhile established and will be reported elsewhere.

## Acknowledgements

Parts of the results reported here were found during a stay at the University of Western Australia, Perth. It is a pleasure to thank Ian McArthur for lots of stimulating discussions during my stay. With equal pleasure I thank Bob Delbourgo for helpful hints and comments, and David Broadhurst for the most exciting and still ongoing collaboration on the subject.

This work was supported under grant A69231484 from the Australian Research council.